\documentclass[pra,twocolumn,superscriptaddress,aps,longbibliography]{revtex4-2}
\usepackage[english]{babel}
\usepackage[dvips]{graphicx}
\usepackage{latexsym}
\usepackage{amsthm}
\usepackage{amsfonts}
\usepackage{amssymb}

\usepackage[colorlinks=true,citecolor=blue,linkcolor=red,urlcolor=blue, linktocpage=true,breaklinks=true,pagebackref=false]{hyperref}

\usepackage{textcomp} 		
\usepackage{hyperref} 		
\usepackage{ifthen} 			
\usepackage{xifthen} 		
\usepackage{color} 			
\usepackage{soul} 			
\usepackage{amsmath} 		
\usepackage{lineno}
\usepackage[normalem]{ulem}

\usepackage{xcolor} 
\usepackage{float}

\newcommand{\uproman}[1]{\uppercase\expandafter{\romannumeral#1}}

\begin{document}


\title{Amplification and spectral evidence of squeezing in the response of a strongly driven nanoresonator to a probe field
}
\author{J.\,S. Ochs}
\thanks{Formerly Huber}
\affiliation{Department of Physics, University of Konstanz, 78457 Konstanz, Germany}
\author{M. Seitner}
\affiliation{Department of Physics, University of Konstanz, 78457 Konstanz, Germany}
\author{M.\,I. Dykman}
\affiliation{Michigan State University, East Lansing, MI 48824, USA}
\email{dykmanm@msu.edu}
\author{E.\,M. Weig}
\affiliation{Department of Physics, University of Konstanz, 78457 Konstanz, Germany}
\affiliation{Present Address: Department of Electrical and Computer Engineering, Technical University of Munich, 80333 Munich, Germany}
\email{eva.weig@tum.de}

\begin{abstract} 

Because of their small decay rates, nanomechanical modes enable studying strongly nonlinear phenomena for a moderately strong resonant driving. Here we study the response of a driven resonator to an additional probe field. We experimentally demonstrate resonant amplification and resonant absorption of the probe field. The corresponding spectral peaks lie on the opposite sides of the strong-drive frequency. Even though the fluctuation-dissipation theorem does not apply, we show that the response to the probe field allows us to characterize the squeezing of fluctuations about the stable states of forced oscillations. Our two-tone experiment  is done in the classical regime, but our findings should equally apply to quantum fluctuations as well. In quantum terms, the observed response is due to multiphoton processes. The squeezing parameter extracted from the spectra of the response is in excellent agreement with the calculated value with no free parameters.   

\end{abstract}

\date{\today}

\maketitle

\section{Introduction} 
Nanomechanical vibrational systems have been traditionally studied in the context of mesoscopic condensed matter physics and nonlinear dynamics,  and more recently also from the perspective of quantum and classical nonlinear optics \cite{Cleland2003,Dykman2012,Schmid2016} and quantum information \cite{O'Connell2010,Satzinger2018,Chu2018}. Different nanomechanical vibrational modes are strongly localized within the resonator, they are well separated in frequency and well-characterized. They often have a very high quality factor $Q$, which is determined by the ratio of the vibration frequency to the decay rate and can exceed $10^8$ \cite{Tsaturyan2017,Ghadimi2018}. The modes can be easily driven into a nonlinear regime, and a host of nonlinear phenomena has been explored, from vibration bistability and associated effects,  cf.~\cite{Buks2001,Postma2005,Aldridge2005,Stambaugh2006,Kacem2009,Unterreithmeier2010,Defoort2015,Dolleman2019}, to nonlinear mode coupling, frequency mixing, chaos, and a frequency comb generation,  cf.~\cite{Erbe2000,Almog2006,Karabalin2009,Westra2010,Eichler2012,Antoni2012,Castellanos-Gomez2012,Defoort2013,Mahboob2015,Guttinger2017,Seitner2017,Czaplewski2018}, and to processes that involve spontaneous or stimulated Raman scattering and are largely exploited in the context of cavity optomechanics~\cite{Aspelmeyer2014}.  

An important group of applications of nanomechanical systems is related to the mass, charge, and force sensing. A fundamental constraint on the sensitivity of a measurement is imposed by noise. The noise comes from classical sources as well as from quantum fluctuations. A by now well-established by now approach to suppressing quantum noise is based on using squeezed states \cite{Caves1981}. It has been implemented recently in laser interferometers for gravitational wave detection \cite{Tse2019,Acernese2019}. With the techniques of cavity optomechanics, squeezing in the quantum regime has been achieved also in mechanical systems \cite{Wollman2015,Lecocq2015,Pirkkalainen2015a}. In quantum squeezing, the variance of one of the quadratures of the vibrations is reduced below its value in the ground vibrational state, whereas fluctuations of the other quadrature are increased. The reduction and the increase come together as a consequence of the uncertainty principle, since the quadrature operators correspond to the scaled vibration coordinate and momentum in the rotating frame and do not commute. 

However, squeezing is not limited to the quantum domain. One may expect squeezing to occur for classical fluctuations, too, in which case fluctuations of one of the quadratures are smaller than in thermal equilibrium. For a degenerate parametric mechanical amplifier, squeezing of classical fluctuations was first demonstrated by Rugar and Gr\"utter \cite{Rugar1991}. 

Squeezing should  generically emerge in periodically driven vibrational systems. This is a consequence of the broken continuous time-translation symmetry.  Indeed, the quadratures are the vibration components that oscillate as $\cos\omega_0t$ and $\sin\omega_0t$, where $\omega_0$ is the vibration frequency. If the system has a continuous symmetry, the origin of time can be shifted.  A shift of $\pi/2\omega_0$ results in the interchange of the quadratures, which shows that the variances should be equal.  A periodically driven system, in contrast, has a discrete time-translation symmetry. It is symmetric only with respect to changing time by the period of the drive.  Therefore the quadratures may no longer be interchanged and their variances are generally different.  

It follows from the above argument that, along with the traditionally explored squeezing due to parametric driving \cite{Walls2008}, one may expect squeezing of fluctuations in a driven nonlinear vibrational mode, and for resonant driving the effect may be resonantly strong. The occurrence of squeezing in this case could be inferred  from the early work on resonantly driven nonlinear modes \cite{Dykman1979b,Dmitriev1986}. A theory of squeezing was developed by Buks and Yurke \cite{Buks2006a}, and a strong suppression of a spectral component of one of the quadrature was first observed in a  nanomechanical Duffing resonator by Almog et al. \cite{Almog2007a}. This observation was based on the conventional homodyne detection scheme and was done in a narrow parameter range near the cusp on the bifurcation curve.

Homodyne measurements are strongly impeded by frequency fluctuations, which play an important role in nanomechanical systems. The limitations are particularly strong in systems with small damping, where the uncertainty in the in-phase component due to slow frequency fluctuations becomes large \cite{Fong2012}. No homodyne measurements of squeezing have been reported for strongly underdamped vibrational systems, to the best of our knowledge. However, it was demonstrated \cite{Huber2020}, that the squeezing of classical fluctuations in underdamped resonantly driven systems can be found by measuring their power spectrum, which is profoundly asymmetric.  The asymmetry comes along with the squeezing \cite{Dykman2012a} as a consequence of a resonant driving, which allows one to find the squeezing parameter in the classical regime.

In this paper we demonstrate that the squeezing parameter of a strongly driven underdamped nanomechanical system can also be determined by measuring the spectrum of the response of the system to an additional weak probe force. A major advantageous feature of this result is that, although the experiment is carried out where the dynamics of the system is classical, the method is not limited to the classical regime which is addressed by our experiment. It can be equally well applied to characterize squeezing of quantum fluctuations in a strong resonant field, since the features of the response to a weak field are temperature-independent.  

For weak damping the absorption spectrum of the probe field should display two peaks, where the mode has one stable state of forced vibrations and up to five peaks where there are two stable vibrational states \cite{Dykman1979b,Dykman2012a,Dykman1994c}. Expanding on the previous results, we show that the difference between the two peaks is determined by the squeezing parameter, which is similar to the case of the power spectrum in the classical regime \cite{Huber2020}.  The onset of the two peaks induced by a probe force in the vibration amplitude of a mechanical mode was seen by  Antoni {\it et al.} \cite{Antoni2012}, but the question of squeezing was not addressed  in this paper.

Besides the squeezing, we wish to highlight another aspect of the response of the resonantly driven system to an additional
 probe force. One of the two peaks corresponds to an amplification of the corresponding field.
 \footnote{Both the strong drive and the probe force come from electromagnetic fields. Therefore, we are using the term field whenever we want to emphasize the physical source of the force and the energy aspect of the driving.} 
 In other terms, rather than taking energy from the field, the mode gives energy to it. Such amplification can be thought of in terms of parametric amplification in nonlinear optics, where a probe light beam is amplified by a strong beam \cite{Shen1984}. 	However, in the present case the process is stongly nonlinear, and can not be described as a regular four-wave mixing for which the response of a driven nonlinear mode to a probe field \cite{Dykman1979b,Dykman1994c} can be thought of as a combination of the ``signal'' and ``idler'' components. In a micromechanical resonator these components were observed in Ref.~\cite{Almog2006}. Rather, the amplification discussed here is a multiphoton process, in optics terms, as we explain in Sec.~\ref{sec:MD_theory}. We report a direct  observation of this effect in our classical setting.

We describe in Sec.~\ref{sec:setup} the setup of the experiment and in Sec.~\ref{sec:observations}  the experimental observation of the two-peak response spectrum, with one of the peaks corresponding to the amplification of the probe drive.  In Sec.~\ref{sec:MD_theory} we briefly overview fluctuations about the stable states of forced vibrations of a weakly damped mode and introduce the squeezing parameter.  Section \ref{sec:susceptibility_theory} gives explicit expressions of the susceptibility in terms of this parameter. Section~\ref{sec:exact_resonance} describes the results of the measurements of the squeezing parameter and its dependence on the driving force using the response to the weak probe field. Section~\ref{sec:conclusions} contains a summary.

\section{Setup and Characterization}
\label{sec:setup}

\begin{figure}[t!]
	\centering
	\includegraphics[width=1\linewidth]{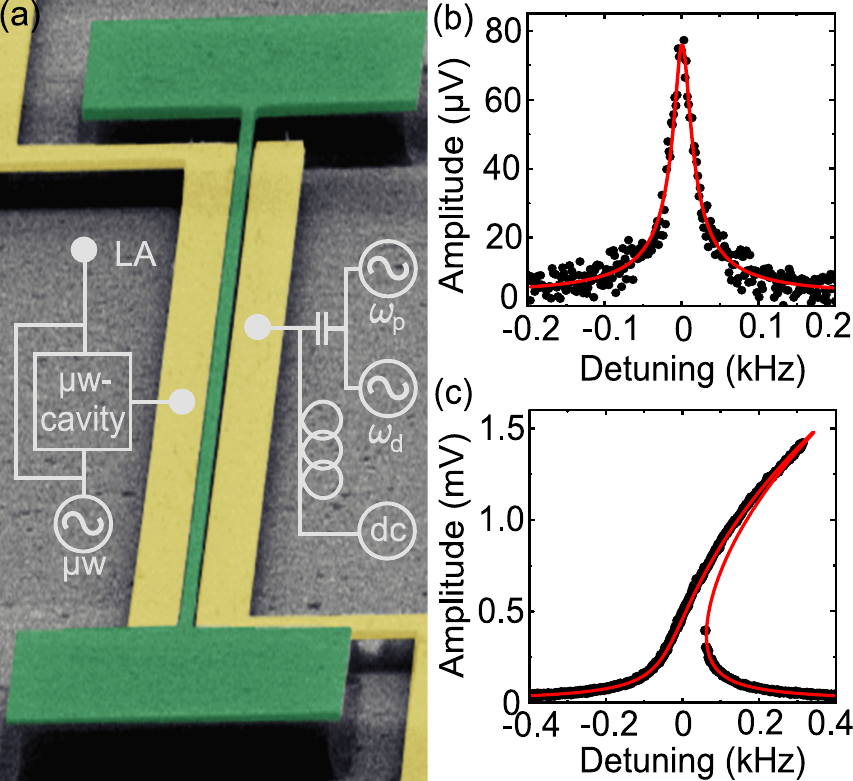}
	\caption{\textbf{(a)} Scanning electron micrograph of the doubly clamped silicon nitride string resonator (green) and two adjacent gold electrodes (yellow) for dielectric drive and detection. The inset displays a schematic of the experimental setup.
	\textbf{(b)} Linear response of the fundamental flexural out-of-plane mode at a drive voltage of $V_{d} = 1$\,mV (black dots). A Lorentzian fit (red solid line) yields an eigenfrequency of 6.528\,MHz, a linewidth of $2\Gamma /2 \pi$ = 20\,Hz, and a quality factor of approximately 325,000.
	\textbf{(c)} Duffing response at a drive voltage of $V_{d} = 20$\,mV (black dots) and fit to the Duffing model (red). The resulting Duffing nonlinearity is $\gamma /(2\pi)^2 = 2.8 \cdot 10^{15} \rm{V}^{-2}\rm{s}^{-2}$. 
}
	\label{fig:fig1}
\end{figure}

The nanomechanical resonator under investigation is a doubly clamped, strongly pre-stressed silicon nitride string resonator fabricated on a fused silica substrate, similar to the one depicted in Fig.\ref{fig:fig1}\,(a) and described in reference \cite{Huber2020}. It is $270$\,nm wide, $100$\,nm thick and $55\,\rm \mu$m long. The string resonator (green) is flanked by two adjacent gold electrodes (yellow), enabling the dielectric transduction combined with a microwave cavity-enhanced heterodyne detection scheme discussed in \cite{Unterreithmeier_2009,Faust_2012,Rieger_2012}. A schematic circuitry of the setup is displayed in the inset of Fig.\ref{fig:fig1}\,(a). The microwave cavity is pumped on resonance at approximately$ 3.6$\,GHz to facilitate displacement detection while avoiding unwanted dynamical backaction effects. 
Actuation and eigenfrequency tuning of the string is accomplished by applying a dc voltage along with a near-resonant rf drive voltage $V_{d}\cos(\omega_{d}t)$. A weak probe tone $V_{p}\cos(\omega_{p}t)$ is additionally scanned across the resonance to record the response spectrum of the device within a small frequency span. 
The two-tone measurements are performed using a fast lock-in amplifier (LA) with a multi-frequency option.
For all measurements presented in the following, a constant dc voltage of $5$\,V is applied such that the fundamental flexural out-of-plane mode can be considered independently, avoiding a hybridization with the in-plane mode \cite{Faust_2013}. The experiment is performed at room temperature of $293$\,K and under vacuum at a pressure below $ 10^{-4}$\, mbar.

The displacement $q(t)$ of the single mode can be described by the equation of motion

\begin{align}
\label{eq:Duffing2Tone}
\ddot{q} + &2 \Gamma \dot{q} + \omega_0^2 q + \gamma q^3 = \nonumber\\
& F_{d} \cos(\omega_{d} t) +F_p \cos(\omega_p t) +\xi(t)\, .
\end{align}

Here, $\omega_0=2\pi f_{0}$ is the angular eigenfrequency, $\Gamma$ the damping rate and $\gamma$ the Duffing nonlinearity parameter.
A comparatively strong driving force with amplitude $F_{d}\propto V_d$ and frequency $\omega_{d} = 2 \pi f_{d}$ is applied; in this experiment, 
we use a resonant drive and set $\omega_{d} = \omega_0$. The additionally applied weaker force probing the response of the device in the vicinity of the resonance has amplitude $F_{p}\propto V_p$ and frequency $\omega_{p} = 2 \pi f_{p}$. The third force term $\xi(t)$ represents the thermal noise. 
The effective mass of the resonator is set to $m = 1$, for the time being.
As both the drive tones and the measured signal are voltage signals, we calibrate the system in units of volts, as discussed in detail in the Sup. Mat. of Ref.~\cite{Huber2020}. 

In the absence of the weak probe tone ($V_{p} = 0$) and using a small amplitude of the drive $V_{d}$ we sweep the frequency $\omega_{d}$ to study the linear response of the fundamental out-of-plane mode at an eigenfrequency of $f_0 = 6.528$\,MHz. It is shown for a drive of $V_{d} = 1$\,mV as a function of the detuning $f_{d}-f_0$ in Fig.\ref{fig:fig1}\,(b) as black dots. A Lorentzian fit (red solid line) yields a linewidth $2\Gamma/2\pi = 20$\,Hz and a quality factor of $Q\approx 325,000$. 

Increasing the drive voltage leads to the well-known Duffing response. A bidirectional response curve at a drive voltage of $V_{d} = 20$\,mV is plotted in Fig.\ref{fig:fig1}\,(c) as black dots. A fit to the Duffing model (red line) allowed us to extract the Duffing nonlinearity parameter $\gamma /(2\pi)^2 = 2.8 \cdot 10^{15} \rm{V}^{-2}\rm{s}^{-2}$.

\section{The linear response spectrum and the amplification of the probe field}
\label{sec:observations}

\begin{figure}[t!]
	\centering
	\includegraphics[width=1\linewidth]{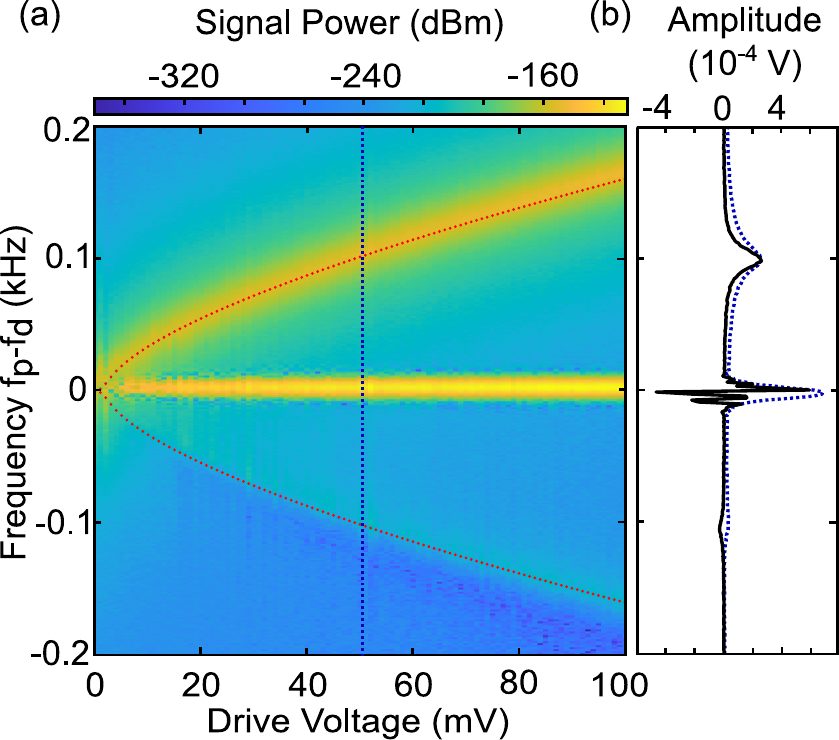}
	\caption{\textbf{(a)} Color-coded response spectra for increasing drive voltages at $f_{d} = f_{0}$. The probe tone is fixed to $V_{p}=3$\,mV. A central band and two satellite peaks are clearly visible. The satellites have strongly different brightness. Their splitting increases with the increasing amplitude of the drive, in good agreement with the theoretical model of Eq.~(\ref{eq:weak_damping}) which is plotted as red dots. The blue dotted line indicates the line cut discussed in (b). 
	\textbf{(b)} Amplitude (blue dotted line) and quadrature component (black solid line) of the vibrations at the probe frequency for $V_{d} = 50$\,mV.}
	\label{fig:figure2}
\end{figure}

The measurements described in the following are performed using the two-tone scheme introduced in Fig.~\ref{fig:fig1}(a). The resonant drive tone applied at $f_{d}=f_0$ is increased from $0$ to $100$\,mV. 
As the drive is applied on resonance, the nonlinear resonator exhibits only one stable state. The weak probe tone with amplitude $V_{p}$ is swept across the resonance with a bandwidth of $5$\,Hz and an eighth-order filter of the fast lock-in amplifier. We record the in-phase and quadrature components of the vibrations at the probe frequency. We verified that the amplitude of these vibrations is proportional to $V_p$ in the studied range of $V_p$, indicating  that the results refer to the regime of the linear response to the probe force.  Figure~\ref{fig:figure2}(a) displays the spectra of this response to a probe of $V_{p} = 3$\,mV for different amplitudes of the strong force  $V_{d}$ using a color-coded amplitude.

Two distinct peaks, equally spaced from the line at the frequency of the strong drive (i.e., ${f}_{p}-f_{d}=0$) and with a power-dependent splitting are observed. As will be discussed in Sec.~\ref{sec:susceptibility_theory}, these two peaks originate from the probe-induced small-amplitude vibrations about the stable state of the vibrations induced by the strong drive. The bright horizontal band centered at the strong-drive frequency indicates the response of the resonator to the probe interfering with the strong drive tone. 

The blue dotted line in Fig.\,\ref{fig:figure2}\,(a) indicates a single line scan of the response to the probe. It refers to a driving voltage of $V_{d} = 50$\,mV. The amplitude of the vibrations at the probe frequency $\omega_p$ is shown as a blue dotted line in Fig.\,\ref{fig:figure2}\,(b). In addition, the quadrature component of these vibrations is plotted as a black solid line. These will be the signal components which will be discussed in more detail below. In the  both data sets the higher frequency satellite is much brighter than that at the lower frequency. As we will also discuss in the following, this is a spectral evidence of the thermomechanical squeezing induced by the strong drive, similar to what has been reported for the power spectrum \cite{Huber2020}. In contrast to the amplitude data which features two positive satellites, the quadrature data exhibits different signs for the two satellites, a positive higher-frequency satellite, and a negative lower-frequency satellite.

Both satellites in the quadrature data have a Lorentzian line shape with the same linewidth 2$\Gamma/2\pi =20$\,Hz as the resonator. In contrast, the dependence of the amplitude on $\omega_p$ is not Lorentzian, whereas the peaks of the \textit{squared} 
amplitude again have a Lorentzian shape (see Sec.~\ref{sec:susceptibility_theory}).

\section{Squeezing of underdamped vibrations in the rotating frame}
\label{sec:MD_theory}

Prior to analyzing the manifestation of squeezing in the response to the probe force, we consider the dynamics in the absence of the probe force and noise, i.e., for $F_p=0$  and $\xi(t)=0$ in Eq.~(\ref{eq:Duffing2Tone}). 
 This  extends the results of Ref.~\onlinecite{Huber2020} to the semiclassical domain for the sake of a more general analysis of the susceptibility in the following sections.\\
 The driving force in this case is  just sinusoidal, $F_d\cos\omega_dt$. The stable states of forced vibrations of a mode are formed by the balance of the energy absorption from the driving force  and the energy drain to a thermal reservoir. The underlying process is strongly nonlinear.  The mode frequency depends on the amplitude, and thus, by changing the vibration amplitude, the driving field ``prepares'' the absorption coefficient. For the considered resonant field, weakly damped modes display strong nonlinearity  when the field is comparatively weak. The forced vibrations   are essentially sinusoidal, but their amplitudes (and phases) can take two values, in a certain parameter range. The stable  values of the squared amplitude $A^2_1$ and $A^2_2$  are determined, respectively, by the largest and smallest roots of the equation \cite{Landau2004a} 
\begin{align}
\label{eq:amplitudes}
&\varphi(\rho_j)=0,  \quad \rho_j = 3|\gamma|A_j^2/8\omega_{d}\Gamma,\\
&\varphi(\rho) = \rho[(\rho-\Omega\, \mathrm{sgn}\gamma)^2+1]-\frac{3|\gamma| F_{d}^2}{32\omega_{d}^3\Gamma^3}, \quad \Omega = \frac{\omega_{d} -\omega_0}{\Gamma}.\nonumber
\end{align}
In what follows we assume $\gamma >0$; this condition holds in our system. The occurrence of two stable values of $A_j$ is seen in Fig.~\ref{fig:fig1}(c). The higher and lower branches of $A_j$ as a function of $\omega_{d}$ in this figure correspond to $j=1$ and $j=2$ in Eq.~(\ref{eq:amplitudes}), respectively. 

For weak damping and weak nonlinearity it is convenient to describe the dynamics of the mode by switching to the rotating frame at the drive frequency $\omega_{d}$. We introduce the scaled coordinate and momentum, $Q$ and $P$, respectively, which correspond to the in-phase and the quadrature components of the vibrations, 
\[q(t) +i\omega_{d}^{-1} p(t) = (Q +iP)\exp(-i\omega_{d}t), \qquad p(t)\equiv \dot q(t) \]
[we note that the variables $Q$ and $P$ differ from the dimensionless variables $Q$ and $P$ in the previous work of one of us (MD), cf. \cite{Dykman2012a}, by a factor 
$(8\omega_{d}\Gamma\Omega/3\gamma)^{1/2}$].
The equations of motion for $Q(t)$ and $P(t)$ follow from Eq.~(\ref{eq:Duffing2Tone}). If we disregard small fast-oscillating corrections, i.e., use the rotating wave approximation \cite{Landau2004a}, the coefficients in these equations are independent of time. A $j$th stable state corresponds to a stable stationary solution $(Q_j, P_j)$ of these equations, and $Q_j^2 + P_j^2 = A_j^2$. 

The dynamics near a stable state is described by linearizing the equations of motion in $Q-Q_j$ and $P-P_j$. In the weak-damping limit the trajectories $Q(t)-Q_j$ and $P(t)-P_j$ correspond to weakly decaying oscillations. The oscillation frequency is
\begin{align}
\label{eq:weak_damping}
\omega_j = \Gamma [(3\rho_j-\Omega)(\rho_j-\Omega)]^{1/2}.
\end{align}
The weak-damping condition that we refer to is
\begin{align}
\label{eq:weak_damping_define}
\omega_j\gg \Gamma.
\end{align}
 We emphasize that $\omega_j$ is the oscillation frequency in the rotating frame,  $\omega_j\ll \omega_0$. Therefore the inequality $\Gamma\ll \omega_j$ is a much stronger constraint on the decay rate $\Gamma$ than the condition $\Gamma\ll \omega_0$ that the mode is underdamped. In our experiment the condition $\Gamma\ll \omega_j$ was satisfied except for a narrow range of small drive amplitudes $F_d$. We note that, in the weak-damping limit Eq.~(\ref{eq:weak_damping_define}), $A_j^2\approx Q_j^2$ and thus $\rho_j\approx 3|\gamma| Q_j^2/8\omega_{d}\Gamma$.

In the presence of a weak noise the driven mode mostly performs small-amplitude fluctuations about the stable state $j$ it occupies. The power spectrum of these fluctuations (measured in the laboratory frame) has peaks at the frequencies $\omega_{d}\pm\omega_j$. Such well-resolved peaks were seen in our previous experiment \cite{Huber2020}. The experiment was done in the classical regime, $k_BT\gg \hbar\omega_0$, and it was found that the areas of the peaks are different. Moreover, the area of one of the peaks was smaller than the area of the peak in the power spectrum at frequency $\omega_0$ in the absence of the driving as given by the fluctuation-dissipation theorem. This is a clear demonstration of the effect of squeezing of  thermal fluctuations.

The squeezing is immediately seen in the expressions for the variances of the in-phase and quadrature components. With account taken of both classical and quantum fluctuations \cite{Huber2020,Dykman2012a}
\begin{align}
\label{eq:variances}
&\langle \delta Q_j^2\rangle \equiv \langle (Q - Q_j)^2\rangle = \frac{\hbar }{4\omega_{d}}(2\bar n +1)(1+e^{-4\phi_j}), \nonumber\\
&\langle \delta P_j^2\rangle \equiv\langle (P-P_j)^2\rangle = \frac{\hbar}{4\omega_{d}} (2\bar n +1)(1+e^{4\phi_j}),
\end{align}
where $\bar n$ is the Planck number of the mode, $\bar n = [\exp(\hbar\omega_0/k_BT) -1]^{-1}$. The squeezing parameter $\phi_j$ is given by the equation
\begin{align}
\label{eq:phi}
\tanh\phi_j = \frac{|3\rho_j-\Omega)|^{1/2} - |\rho_j-\Omega|^{1/2}}{|3\rho_j-\Omega)|^{1/2} + |\rho_j-\Omega|^{1/2}}.
\end{align}
The sign of the parameter $\phi_j$ is determined by the sign of $\rho_j-\Omega$ (which coincides with the sign of $3\rho_j -\Omega$). One can see from Eqs.~(\ref{eq:amplitudes}) and (\ref{eq:weak_damping}) that, on the high-amplitude branch in Fig.~\ref{fig:fig1}(c), i.e., for $j=1$, we have $\phi_1 >0$. This means that the fluctuations of the in-phase component $Q$ are squeezed, whereas those of the quadrature component $P$ are enhanced. For the lower branch, $j=2$, the squeezing parameter is negative, $\phi_2<0$.

As seen from Eqs.~\eqref{eq:amplitudes} and \eqref{eq:phi}, in the considered weak-damping limit, the squeezing parameter depends on a single combination of the parameters of the driven mode 
\begin{align}
\label{eq:beta}
\beta = 3\gamma F_{d}^2/[32\omega_{d}^3(\omega_{d}-\omega_0)^3] \quad (\gamma >0).
\end{align}
As indicated in Eq.~(\ref{eq:beta}), the expression for $\beta$ is written for the considered case $\gamma>0$; an extension to the case $\gamma<0$ is straightforward. The parameter $\beta$ characterizes the strength of the drive. It is proportional to the squared drive amplitude scaled by the cube of the detuning of the drive frequency from the mode eigenfrequency. It is particularly convenient for studying the mode dynamics for weak damping, where the vibration frequencies $\omega_{1,2}$ are high compared to the decay rate \cite{Dykman1979b}. 

\begin{figure}
\includegraphics[width=1\linewidth]{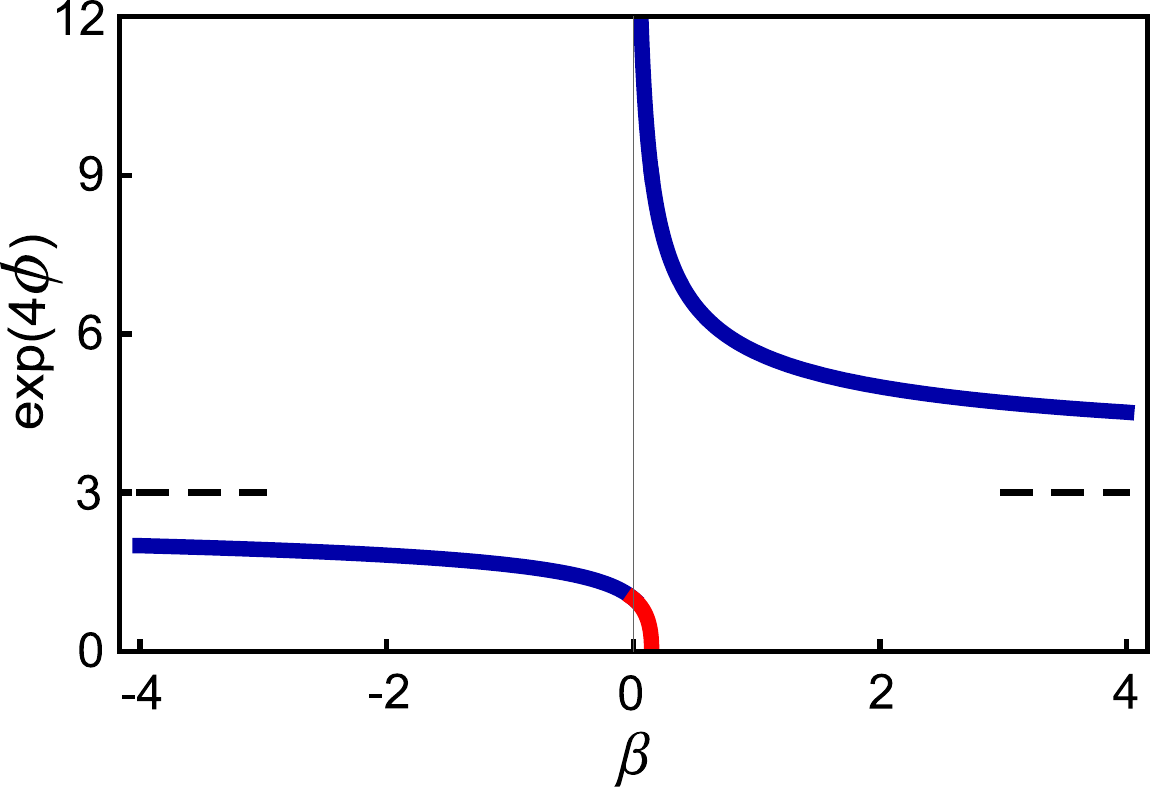}
\caption{Dependence of the squeezing parameter $\phi$ on  the dimensionless strength of the drive $\beta$, Eq.~(\ref{eq:beta}), in the limit of weak damping  and for $\gamma>0$. In this limit ($\omega_1\gg \Gamma$ and, where applicable, $\omega_2\gg \Gamma$)  the bistability of forced vibrations, and thus the vibrational branch $j=2$ exist in the range $0<\beta <4/27$. The red line shows $\exp(4\phi_2)$ in this range. The blue lines show $\exp(4\phi_1)$. Outside the bistability region the system has only one stable vibrational state, and $\phi_1\equiv\phi$. The dashed lines show the limit of $\exp(4\phi_1)$ for $|\beta|\to \infty$ which, physically, corresponds to the driving frequency approaching the mode eigenfrequency, $|\omega_{d}-\omega_0|\to 0$. The weak-damping limit breaks down for small $\beta \to 0$ and, for the branch $j=2$, for $\beta$ approaching $4/27$.}
\label{fig:phi}
\end{figure}

The squeezing parameter $\phi$ as a function of $\beta$ is shown in Fig.~\ref{fig:phi}; in fact, we show directly the relevant parameter $\exp(4\phi)$ (see Eq.~\ref{eq:variances}). As seen from Eqs.~(\ref{eq:weak_damping}) and (\ref{eq:phi}) and from Fig.~\ref{fig:phi}, $\phi$ monotonically decreases to zero as $\beta$ increases from $-\infty$ to 0 (i.e., $|\beta|$ decreases from $\infty$ to 0). In this range of $\beta$ the driven mode has only one stable vibrational state, and the fluctuations are less squeezed as the vibration amplitude decreases with the decreasing $|\beta|$. For $\beta <0$ and $|\beta|\ll 1$ we have $\exp(4\phi_1)\approx 1 +2|\beta|$; the constraint on the applicability of this expression imposed by the weak-damping condition $\omega_1\gg \Gamma$  is $|\Omega|\gg 1$. 

In the range of bistability, $0<\beta<4/27$, both $\phi_1$ and $\phi_2$  decrease with the increasing $\beta$. The function $\exp(4\phi_1)$ is $\approx 2/\sqrt{\beta}$ for small $\beta >0$. This large value of $\phi_1$ for small $\beta>0$ is a consequence of the strong asymmetry of the  phase portrait near the states on branch 1 \cite{Dykman1979b}. The weak-damping constraint on $\beta$ from below in this case is $\Omega\beta^{1/4}\gg 1$. For the branch 2, close to the weak-damping bifurcation point $\beta_B=4/27$, we have $\exp(4\phi_2)  \approx 9\sqrt{(\beta_B-\beta)/4}$. The weak-damping condition in this region is $\Omega (\beta_B-\beta)^{1/4}\gg 1$.

It follows from Eq.~(\ref{eq:variances}) that the variance of the in-phase component is below the standard limit $(\hbar/2\omega_0)(2\bar n+1)$ for an undriven mode, which indicates the squeezing. At the same time, the product of the root-mean-square deviations $[\langle(Q-Q_j)^2\rangle\langle (P-P_j)^2\rangle]^{1/2} = (\hbar/2\omega_{d})(2\bar n +1)[(1+\cosh 4\phi_j)/2]^{1/2}$ is larger than this product for an undriven mode. This indicates that, overall, {\it the driving enhances fluctuations}. Such an increase can be interpreted as an efficient heating of the mode at the expense of the energy absorbed from the strong driving field.

\section{The effect of squeezing on the susceptibility spectrum of a strongly driven mode}
\label{sec:susceptibility_theory}

The response of a driven mode to an additionally applied weak probe force $F_p\cos(\omega_p t)$  can be described by the linear response theory \cite{Dykman1979b,Dykman1994c}. We are interested in the case of resonant driving, where the frequencies of both the strong drive and the weak probe are close to the mode eigenfrequency, $|\omega_p-\omega_0|, |\omega_{d}-\omega_0|\ll \omega_0$. For the mode localized near a $j$th stable state, the resonant increment in the displacement, which is linear in the probe field, has the form
\begin{align}
\label{eq:suscept_definition}
\langle \delta q_j(t)\rangle =& \frac{1}{2}F_p\left[\chi_j(\omega_p)e^{-i\omega_pt} + \mathcal{X}_j(\omega_p)
e^{i(2\omega_{d}-\omega_p)t}\right]\nonumber\\
& + \mathrm{c.c.}
 \end{align}
Generally, there is also a characteristic term that comes from the probe-field induced change of the rates of fluctuation-induced switching between the states $j=1,2$ \cite{Dykman1979b,Dykman1994c}; in micromechanical systems the effect of this term was studied by Stambaugh and Chan \cite{Stambaugh2006a}.  
 
The susceptibility $\chi_j(\omega_p)$ describes the response at the probe frequency $\omega_p$ (the signal component, in the optics terms). In contrast, the susceptibility $\mathcal{X}_j$ shows the occurrence of the probe-induced vibrations at the combination frequency $2\omega_{d}-\omega_p$ (the idler component, in the optics terms). Equation (\ref{eq:suscept_definition}) has the form of the nonlinear susceptibility that describes four-wave mixing, which involves two photons of the strong drive and one photon of the probe \cite{Shen1984}.  Formally, the higher-order terms that involve more than two photons of the strong field seem to have been disregarded. However, in reality both susceptibilities $\chi_j$ and $\mathcal{X}_j$ are multiphoton: the strong drive prepares the stable state $j$ as a result of a process in which many photons of the strong field. The classical language adequately describes such multiphoton dynamics exactly because it is multiphoton; in quantum terms, the analysis corresponds to the WKB approximation, as indicated below.

\subsection{Resonant features of the susceptibilities}

A clear manifestation of the multiphoton nature of the susceptibilities $\chi_j$ and $\mathcal{X}_j$ in the weak-damping case is that they display sharp resonant structure at the probe field frequencies $\omega_p\approx \omega_{d}\pm\omega_j$. For completeness, the explicit general expressions for the susceptibilities $\chi_j$ and $\mathcal{X}_j$ \cite{Dykman1994c} are reproduced in Eq.~(\ref{eq:susceptibility_general}). Simplifying these expressions for the case of weak damping, $\omega_j\gg \Gamma$, in the region of resonance,  
\[ |\omega_p-\omega_{d} -\mathcal{S}_{} \omega_j|\ll \omega_j,  \qquad \mathcal{S}_{}=\pm 1,\]
and using Eq.~(\ref{eq:phi}), one can write the susceptibilities in the form:
\begin{align}
\label{eq:suscept_explicit}
\chi_j(\omega_p)\approx \frac{i}{4\omega_{d}}\,\frac{1+(-1)^{j-1}\mathcal{S}_{}\cosh 2\phi_j}{\Gamma-i(\omega_p-\omega_{d} -\mathcal{S}_{} \omega_j)},\nonumber\\
\mathcal{X}_j(\omega_p) = \frac{-i}{4\omega_{d}}\,\frac{(-1)^{j-1}\mathcal{S}_{}\sinh 2\phi_j}{\Gamma-i(\omega_p-\omega_{d} -\mathcal{S}_{} \omega_j)}.
\end{align}
Equation  (\ref{eq:suscept_explicit}) expresses both susceptibilities in terms of the squeezing parameter $\phi_j$. It suggests a way of measuring this parameter by measuring the susceptibilities. 

The parameters $|\chi_j|^2$ and $|\mathcal{X}_j|^2$ give the squares of the amplitudes of the probe-field induced vibrations at frequencies $\omega_p$ and $2\omega_{d}-\omega_p$. The dependence of the squared amplitudes on $\omega_p$ is described by the Lorentzian peaks centered at $\omega_p=\omega_{d}\pm\omega_j$,  with halfwidth $\Gamma$. For $\chi_j(\omega)$,  the areas of the peaks 
\[\mathcal{A}_j^{\pm}=\int_{|\omega_p-\omega_{d} \mp \omega_j|\ll \omega_j}|\chi_j(\omega_p)|^2 d\omega_p\]
are 
\begin{align}
\label{eq:peak_areas}
\mathcal{A}_j^{(\pm)} =\frac{\pi}{16\omega_{d}^2}\left[1\pm (-1)^{j-1}\cosh 2\phi_j\right]^2.
\end{align}
The ratio $\mathcal{A}_j^{+}/\mathcal{A}_j^{-}=(\tanh \phi_1)^{-4}$ for $j=1$ and $(\tanh\phi_2)^4$ for $j=2$, which makes it possible to immediately extract the parameter $\phi_j$ from the experiment. 

It is instructive to compare the result of Eq.~(\ref{eq:suscept_explicit}) with the susceptibility of the mode in the absence of strong driving. In the weak-noise limit, this susceptibility has the form $\chi(\omega_p) = (i/2\omega_0)[\Gamma-i(\omega_p-\omega_0)]^{-1} $. This expression coincides with Eq.~(\ref{eq:suscept_explicit}) for $\chi_1(\omega_p)$ if in the latter expression we set $j=\mathcal{S}_{} =1$ (there is only one vibrational branch), $\omega_{d} = \omega_0$, and $\phi_1=0$.

The imaginary part of the susceptibility 
\begin{align}
\label{eq:Im_chi}
\mathrm{Im}~\chi_j(\omega_p)= \frac{\Gamma}{4\omega_{d}}\,\frac{1+(-1)^{j-1}\mathcal{S}_{}\cosh 2\phi_j}{\Gamma^2 +(\omega_p-\omega_{d} -\mathcal{S}_{} \omega_j)^2}
\end{align}
describes the absorption of the probe field by the mode. It displays a Lorentzian peak at $\omega_p= \omega_d + \mathcal{S}_{}\omega_j$, as seen in Fig.~\ref{fig:figure2}(b). In equilibrium it is always positive. However, as seen from Eq.~(\ref{eq:Im_chi}), the peak at frequency $\omega_p=\omega_{d} + \mathcal{S}_{}\omega_j$ is negative  for $(-1)^{j-1}\mathcal{S}_{} <0$ and would be more appropriately called a dip. In other words, for a given branch of the forced vibrations $j$, one of the peaks of Im~$\chi_j(\omega_p)$ is positive and the other is negative.   The negative sign of Im~$\chi_j(\omega_p)$ indicates that the mode is amplifying the probe drive rather than absorbing energy from it.  
The amplification comes at the expense of the strong drive. Overall $\int d\omega_p\,\mathrm{Im}\,\chi_j(\omega_p) = \pi/2\omega_{d}$ is positive, a well-known feature of an oscillator \cite{Perelomov1998}. 

We note that the amplitude of the vibrations at frequency $\omega_p$ is positive by definition. However, where this amplitude is large it does not mean that the mode is amplifying the external drive (the ac voltage, in our case). In particular, it does not mean that, in the presence of an appropriately tuned feedback loop, the mode can generate  an ac voltage. A well-known necessary condition for generation is that the imaginary part of the susceptibility is negative.

The ratio of the areas of the peaks of $|\mathrm{Im}\,\chi_j(\omega_p)|$ at $\omega_p=\omega_{d}\pm\omega_j$ is 
\begin{align}
\label{eq:quadrature_areas}
&\mathcal{Q}_1^{+}/\mathcal{Q}_1^{-} = \tanh^{-2}\phi_1, \quad \mathcal{Q}_2^{+}/\mathcal{Q}_2^{-} = \tanh^2\phi_2,\nonumber\\
&\mathcal{Q}_j^{\pm} =\int_{|\omega_p-\omega_{d} \mp \omega_j|\ll \omega_j}|\mathrm{Im}~\chi_j(\omega_p)| d\omega_p
\end{align}

\subsection{Extension to the quantum domain}
 
In the quantum domain, the theory applies for sufficiently large vibration amplitude where, beside the condition $\omega_j\gg \Gamma$,  the condition that many quantum levels of the mode are occupied, $\omega_{d}A_j^2/\hbar \gg 1$, also holds. The other quantum restriction on the applicability of Eq.~(\ref{eq:suscept_explicit}) is that the nonequidistance of the quasienergy levels (Floquet eigenvalues) of the driven mode, which is a consequence of the nonlinearity, is small compared to dissipative level broadening. To the order of magnitude, it corresponds to $\hbar|\gamma|< \Gamma\omega_{d}^2(2\bar n+1)$; a more exact condition is given in Ref.~\cite{Dykman2012}. Even where this condition does not hold, the expressions for the areas of the spectral peaks Eq.~(\ref{eq:peak_areas}) still apply.

\section{Squeezing for the strong drive tuned to exact resonance}
\label{sec:exact_resonance}

The analytical expressions simplify in the case where the frequency of the strong driving force $\omega_{d}$ is equal to the mode eigenfrequency $\omega_0$. In this case there is only one branch of forced vibrations, $j=1$. As seen from Eqs.~(\ref{eq:phi}) and ~(\ref{eq:quadrature_areas}), the squeezing parameter $\phi$ and the ratio of the peak areas are
\begin{align}
\label{eq:Omega_0}
& \phi\equiv \phi_1 =  (\ln 3)/4, \nonumber \\
& \mathcal{Q}_1^{+}/\mathcal{Q}_1^{-} =7+4\sqrt{3} \approx 13.9 \qquad (\omega_{d}=\omega_0).
\end{align}

The squeezing parameter is independent of the force amplitude $F_{d}$. The corresponding value of $\exp(4\phi)=3$ is shown by the dashed lines in Fig.~\ref{fig:phi}. It is this case that we study in the experiment to demonstrate the efficiency of the method. For the classical mode under investigation, $k_BT\gg \hbar\omega_{d}$, we have from Eq.~(\ref{eq:variances})
 \begin{align}
 \label{eq:class_limit}
& \langle \delta Q^2\rangle \equiv \langle \delta Q_1^2\rangle \approx 2k_BT/3\omega_{d}^2,\nonumber\\
 &\langle \delta P^2\rangle \equiv \langle \delta P_1^2\rangle \approx 2k_BT/\omega_{d}^2.
 \end{align}
For comparison, in thermal equilibrium in the classical limit we have   $\langle \delta Q^2\rangle_\mathrm{eq} = \langle \delta P^2\rangle_\mathrm{eq} = k_BT/\omega_0^2 \approx k_BT/\omega_d^2$.

We note that the independence of $\phi\equiv \phi_1$ from the drive amplitude $F_d$ occurs only for sufficiently strong drive. From Eqs.~(\ref{eq:amplitudes}) and (\ref{eq:weak_damping}), the weak damping condition $\omega_1\gg \Gamma$ for the exact resonance has the form  $\rho_1 \approx (3|\gamma| F_d^2/32\omega_d^3\Gamma^3)^{1/3} \gg 1$, which imposes the constraint on $F_d$ from below.

In order to compare the theoretical model with the experimental data, we add the expected position of the two satellites according to Eq.~(\ref{eq:weak_damping}) as a red dotted line in Fig.\,\ref{fig:figure2}\,(a).  We find excellent agreement between the model (with no free parameters) and the experiment. To further analyze our data, we focus on the imaginary part of the susceptibility. It determines the quadrature component, which is directly determined from the response measurement (see black trace in Fig.~\ref{fig:figure2}(b)).  

Clearly, the different signs of the satellite peaks predicted by the model and attributed to the absorption and amplification of the probe field are recovered. The absorption is observed for the higher frequency satellite with the higher intensity, whereas the amplification is visible in the weaker and lower frequency satellite. We do not measure the power of the probe field at the output, but the very fact that the resonantly absorbed power $(\omega_pF_p^2/2)\text{Im}~\chi_j(\omega_p)$ is negative for Im~$\chi_j(\omega_p)<0$ unambiguously indicates that the nanostring pumps energy into the probe field. 

\begin{figure}[t!]
	\centering
	\includegraphics[width=1\linewidth]{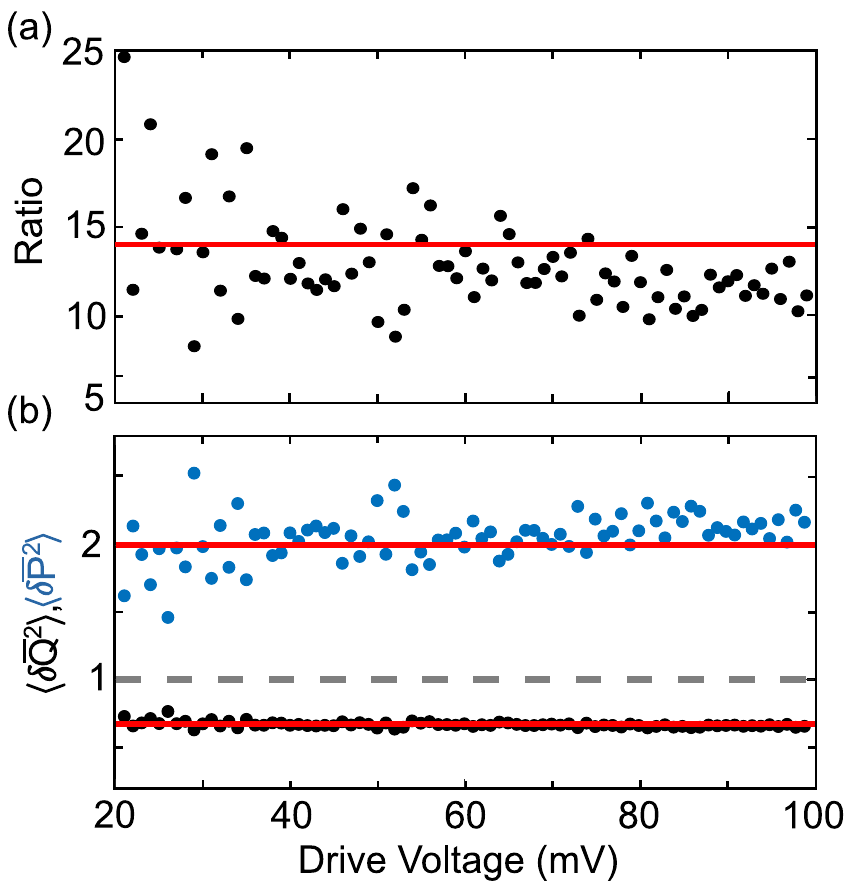}
	\caption{\textbf{(a)} Ratio of the areas of the satellites peaks obtained form the quadrature data as a function of the drive voltage (black dots). Red line corresponds to the theoretical prediction Eq.~(\ref{eq:Omega_0}).
	\textbf{(b)} Normalized dimensionless variances of the in-phase and quadrature fluctuations $\langle \overline{\delta Q^2}\rangle =  (\omega_0^2/k_BT)\langle \delta Q^2\rangle$  and $\langle \overline{\delta P^2}\rangle =  (\omega_0^2/k_BT)\langle \delta P^2\rangle$  around the stable state of the forced vibration as functions of the drive voltage. Black and blue dots show the in-phase and quadrature values extracted from the experimentally determined satellite area ratio from (a), whereas the red lines show the corresponding theoretical model of Eq.~(\ref{eq:class_limit}). The gray dashed line illustrates the thermomechanical fluctuations at the temperature $293$\,K used in the experiment.}
	\label{fig:figure3}
\end{figure}

According to Eq.~(\ref{eq:Im_chi}), the satellites in the imaginary part of the susceptibility $|\mathrm{Im}~\chi_1(\omega_p)|$ have a Lorentzian lineshape. Thus, Lorentzian fits are employed to extract the area enclosed under the satellites, $\mathcal{Q}_1^{+}$ and $\mathcal{Q}_1^{-}$. 
The area ratio $\mathcal{Q}_1^{+}/\mathcal{Q}_1^{-}$ extracted from the imaginary part of the susceptibility is plotted in Fig.~\ref{fig:figure3}(a) as black dots and compared with the theoretical prediction Eq.~(\ref{eq:Omega_0}), which is included as a red line. No free parameters are employed. We find excellent agreement between the experiment and the theory. We note that the amplitude of  the signal, i.e. the blue dotted line in Fig.~\ref{fig:figure2}(b), which is given by $|\chi_j(\omega_p)| \propto [\Gamma^2 + (\omega_p-\omega_{d})^2]^{-1/2}$ is a non-Lorentzian  function of $\omega_p$. It falls off slower than the Lorentzian with the increasing $|\omega_p-\omega_{d}|$, as indeed seen in Fig.~\ref{fig:figure2}~(b). However, $|\chi_j(\omega_p)|^2$ is again fitted by a Lorentzian.

It is expected from Eq.~(\ref{eq:Omega_0}) that the squeezing parameter is independent of the drive amplitude $F_{d}$ if the drive frequency coincides with the eigenfrequency of the mode. Such independence is indeed seen in the experiment, which yields an average squeezing parameter of $\phi \approx 0.28\pm 0.03$ in good agreement with the theoretically obtained value of $\mathrm{arctanh}[1/(7+4\sqrt{3})^{1/2}]\approx 0.27$. This confirms not only the analysis of the squeezing, but also the model we use to describe the mode dynamics.

In Fig.~\ref{fig:figure3}(b), the result is  re-expressed in terms of the mean-square fluctuations of the in-phase and quadrature component $\langle \delta Q^2\rangle$ and $\langle \delta P^2\rangle$, Eq.~(\ref{eq:class_limit}). The experimental data are plotted as black and gray dots, respectively, whereas the theoretical predictions obtained using Eq.~(\ref{eq:class_limit}) are shown as red lines. For the sake of clarity, we chose a normalized dimensionless representation. This obviates reintroducing the effective mass of the resonator, which would otherwise re-appear in the denominator of Eqs.~(\ref{eq:variances}) and (\ref{eq:class_limit}). The grey dashed line indicates the mean-square of the thermomechanical fluctuations at $293$\,K, clearly showing that a significant squeezing of the in-phase component is observed. 

We note that the data spread in Fig.~\ref{fig:figure3}(a) and (b) is larger for weak drive power as a result of an insufficient separation of the satellite peaks. There is also a small systematic tilt arising from a deviation from the Duffing model that comes into play for large drive powers, presumably due to higher-order nonlinearities.

\section{Conclusion}
\label{sec:conclusions}

We have presented in this paper the spectra of the response of a moderately strongly driven nanomechanical mode to a weak probe  drive.  The frequencies of the strong drive and the probe tone are close to the mode eigenfrequency. We found that the response spectra of our very weakly damped mode have a peculiar structure. Both the amplitude of the response and the quadrature component display two pronounced features symmetrically located with respect to the strong-drive frequency. For the considered driving on exact resonance, the features on the high-frequency side are more pronounced than on the low-frequency side. In the case of the amplitude, the features are non-Lorentzian peaks. In contrast, the quadrature component displays a peak on the high-frequency side and a dip on the low-frequency side. Both have a Lorentzian shape with the same halfwidth as the peak in the spectrum of the mode in the absence of strong driving. 

The theory predicts the onset of these features. The distance along the frequency axis between the features and the frequency of the strong drive is expected to be equal to the frequency of oscillations of the strongly driven nonlinear mode about its stable state of forced vibrations in the rotating frame. This frequency, in turn, is determined by the interplay of the strong drive and the mode nonlinearity. Therefore, in terms of quantum optics, the occurrence of the features of the response is a result of multiphoton processes that involve multiple quanta of the strong drive.

The amplitude and the quadrature component of the probe-induced vibrations are determined, respectively, by the absolute value and the imaginary part of the susceptibility of the mode with respect to the probe. The dip in the quadrature spectrum corresponds to a negative imaginary part of the susceptibility. This indicates that the weak probe field is amplified by the mode. The amplification comes at the expense of the energy provided by the strong drive. 

The strongly driven mode is a system away from thermal equilibrium. Therefore there is no standard relation between the imaginary part of the susceptibility and the spectrum of fluctuations of the mode. Nevertheless it is clear on the physical grounds that there should be some relations between these spectra for the considered weakly damped mode. Indeed, in the rotating frame, the stable state of forced vibrations is a stationary state. The weakly damped mode performs random oscillations about this state due to thermal and quantum noise, but these oscillations can also be resonantly excited by an external drive.  Therefore the power spectrum and the susceptibility should display features at the same frequencies. The susceptibility thus encodes some information about the fluctuations. 

The theoretical analysis shows that one of the components of the random oscillations  should be squeezed and that the squeezing determines the ratio of the areas of the dip and the peak in the imaginary part of the susceptibility. Therefore by measuring these areas one can extract the squeezing parameter. Moreover, the theory predicts that, if the frequency of the strong drive is equal to the mode eigenfrequency, the ratio of the areas should be independent of the drive amplitude in a broad amplitude range. This is in excellent agreement with the experiment.  Our findings have been independently reproduced on a second sample, see Appendix~\ref{sec:additional}.  

The agreement between the experiment and the theory regarding the positions of the peaks of the susceptibility, their shape, and their areas, with no adjustable parameters, provides  strong evidence of the squeezing of fluctuations. It complements and significantly extends the results on the shape and the area of the classical power spectra of a driven mode \cite{Huber2020}. The extension is particularly important as the method presented in Ref.~\cite{Huber2020} to extract the squeezing parameter is applicable only in the classical regime, whereas both classical and quantum fluctuations are squeezed in a strongly driven systems. Although the experiment is done in the classical regime, our method does not rely on and does not require that the fluctuations be classical. To the best of our knowledge, reports of an observation of a multiphoton resonant amplification by a driven nonlinear vibrational mode are lacking in the literature. The presented results demonstrate interdisciplinary aspects of the studies of nanomechanical systems.  \\
 Data and analysis code are available at \cite{zenodo}.

\section{Acknowledgements}
J.\,S.\,O. and E.\,M.\,W. gratefully acknowledge financial support from the European Union’s Horizon 2020 Research and Innovation Programme under Grant
Agreement No 732894 (FET Proactive HOT), and the German Federal Ministry of Education and Research
(contract no. 13N14777) within the European QuantERA cofund project QuaSeRT. They also acknowledge support from the Deutsche Forschungsgemeinschaft via collaborative research center SFB 1432.
M.\,I.\,D. acknowledges support from the National Science Foundation, Grants No. DMR-1806473 and CMMI 1661618. M.\,I.\,D. is grateful for warm hospitality at the Zukunftskolleg and the Physics Department of the University of Konstanz where this work was started. The authors thank Gianluca Rastelli and Wolfgang Belzig for fruitful discussions and comments about this work and David Holzapfel for his contributions to reproducing the experimental results on a second sample.

\appendix
\section{ The susceptibility of a strongly driven Duffing oscillator}
 The susceptibilities $\chi_j(\omega)$ and $\mathcal{X}_j(\omega)$ of a driven classical Duffing oscillator in the stable vibrational state $j$ (with $j=1,2$) have the form \cite{Dykman1994c}
 \begin{align}
 	\label{eq:susceptibility_general}
 	&\chi_j(\omega) = \frac{i}{2\omega_d} \frac{\Gamma-i(\omega-\omega_d) -i\Gamma(2\rho_j-\Omega)}
 	{\tilde\omega_j^2-2i\Gamma(\omega-\omega_d) - (\omega-\omega_d)^2},\nonumber\\
 	&\mathcal{X}_j(\omega) = -\frac{\Gamma}{2\omega_d}\frac{\rho_j}
 	{\tilde\omega_j^2-2i\Gamma(\omega-\omega_d) - (\omega-\omega_d)^2},
 \end{align}
 where $\rho_j$ is the scaled squared amplitude of the forced vibrations in the state $j$, which is given by Eq.~(\ref{eq:amplitudes}), and
 \begin{align}
 	\label{eq:nu_j}
 	\tilde\omega_j^2 = \Gamma^2 [1+(3\rho_j-\Omega)(\rho_j-\Omega)]\equiv \Gamma^2+ \omega_j^2.
 \end{align}
 Here $\omega_j$ is the frequency of vibrations about the stable state $j$ in the rotating frame. It is given by Eq.~(\ref{eq:weak_damping}).  In the weak-damping limit defined in Eq.~(\ref{eq:weak_damping_define})
 $\tilde\omega_j \approx \omega_j$. 
 
 The total susceptibility of the driven classical oscillator  is the sum of the partial susceptibilities (\ref{eq:susceptibility_general}) weighted with the populations $w_j$ of the corresponding stable states and the susceptibility $\chi_\mathrm{tr}(\omega)$ related to the modulation of the populations $w_j$ by the probe force. This term  has a characteristic very narrow feature centered at $\omega = \omega_d$ \cite{Dykman1979b,Dykman1994c,Stambaugh2006a}. The expression (\ref{eq:susceptibility_general}) for $\chi_j(\omega)$ was also later given in Ref.~\onlinecite{Defoort2013}. The experimental results in this paper show the dependence of  $\arg\max \sum_jw_j|\chi_j(\omega)|$ on the drive frequency $\omega_d$. The set of equations that give the susceptibilities $\chi_j(\omega)$ and $\mathcal{X}_j(\omega)$ was also given in Ref.~\onlinecite{Almog2006}. We note that the two-peak structure of $\mathrm{Im}\chi_j(\omega)$ is pronounced in the weak-damping limit, $\omega_j\gg \Gamma$. Near a bifurcation point, where $\chi_j$ was considered analytically in Ref.~\onlinecite{Almog2006},  $\mathrm{Im}\chi_j(\omega)$ has a single peak, as $\tilde\omega_j=0$ at a bifurcation point. 
 
 The expressions for the susceptibilities have to be modified in the deeply quantum regime. The corresponding analysis is given in Ref.~\onlinecite{Dykman2012a}.

 \section{ Additional experimental data from sample 2}
\label{sec:additional}

 We repeated the experiment discussed in Secs. \ref{sec:setup}, \ref{sec:observations} and \ref{sec:exact_resonance} on another sample (sample 2) using the same measurement technique as described in Fig.~1(a). For all measurements on the second sample, a constant dc voltage of $10$\,V is applied and the fundamental flexural out-of-plane mode can be considered as a single mode. 
 \subsection{Characterization}
\begin{figure}[H]
 	\centering
 	\includegraphics[width=1\linewidth]{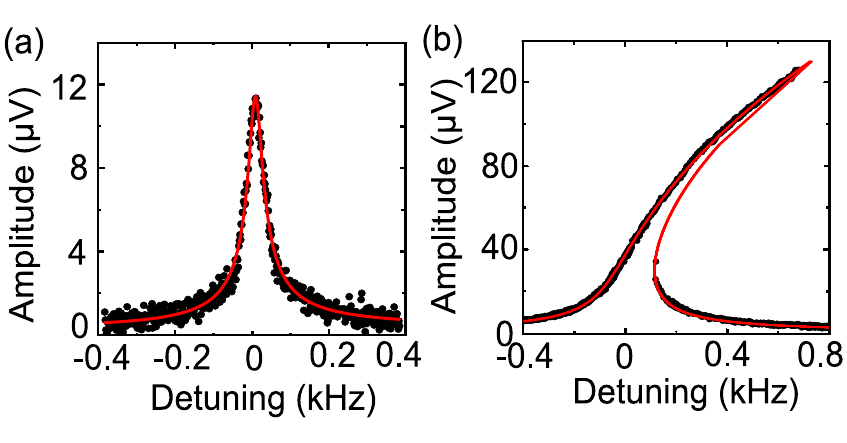}
 	\caption{\textbf{(a)} Linear response of the fundamental flexural out-of-plane mode at a drive voltage of $V_{d} = 3$\,mV (black dots). A Lorentzian fit (red) yields an eigenfrequency of $6.562$\,MHz, a linewidth of $2\Gamma /2 \pi = 35$\,Hz and a quality factor $Q \approx 190,000$.
 		\textbf{(b)} Duffing response at a drive voltage of $V_{d} = 35$\,mV (black dots) and fit with the Duffing model (red). The resulting Duffing nonlinearity is $\gamma /(2\pi)^2 = 7.6 \cdot 10^{17} \rm{V}^{-2}\rm{s}^{-2}$. 
 	}
 	\label{fig:fig1_sm}
 \end{figure}

 The linear response of the fundamental out-of-plane mode is found at an eigenfrequency of $f_{0}=6.562$\,MHz. It is shown for a drive of $V_d$ = 3\,mV as a function of the detuning $f_d-f_0$ in Fig. \ref{fig:fig1_sm}(a) as black dots along with a Lorentzian fit (red solid line). From the fit, we extract a linewidth of $2\Gamma/2\pi = 35$\,Hz and a quality factor of $Q \approx 190,000$.
 A bidirectional response curve in the Duffing regime at a drive voltage of $V_d=$\,35\,mV is plotted in Fig. \ref{fig:fig1_sm}(b). A fit with the Duffing model yields the Duffing nonlinear parameter $\gamma/(2\pi)^2 = 7.6\cdot 10^{17} \rm{V}^{-2}\rm{s}^{-2}$. Details on the fitting procedure are given in the Supplemental Material of Ref.~\cite{Huber2020}. 

 \subsection{Squeezing and Amplification}
\begin{figure}[H]
 	\centering
 	\includegraphics[width=\linewidth]{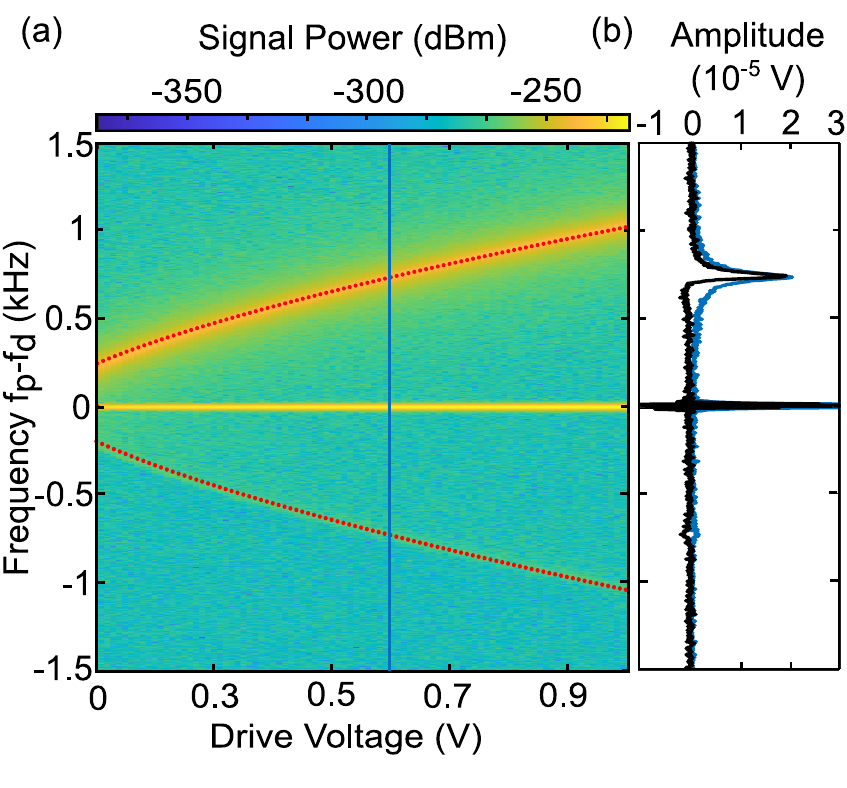}
 	\caption{\textbf{(a)} Color-coded response spectra for increasing drive voltages at $f_{d} = f_{0}$. The probe tone is fixed to $V_{p}=5$\,mV. A central feature and two satellite peaks are clearly visible. The satellites have strongly different brightness. Their splitting increases with the increasing amplitude of the drive, in good agreement with the theoretical model of Eq.\,(3), which is plotted as red dots. The blue  line indicates the line cut discussed in panel (b). 
 		\textbf{(b)} Amplitude (blue line) and quadrature component (black solid line) of the vibrations at the probe frequency for $V_{d} = 0.6$\,V.}
 	\label{fig:figure2_sm}
 \end{figure}

\begin{figure}[H]
 	\centering
 	\includegraphics[width=\linewidth]{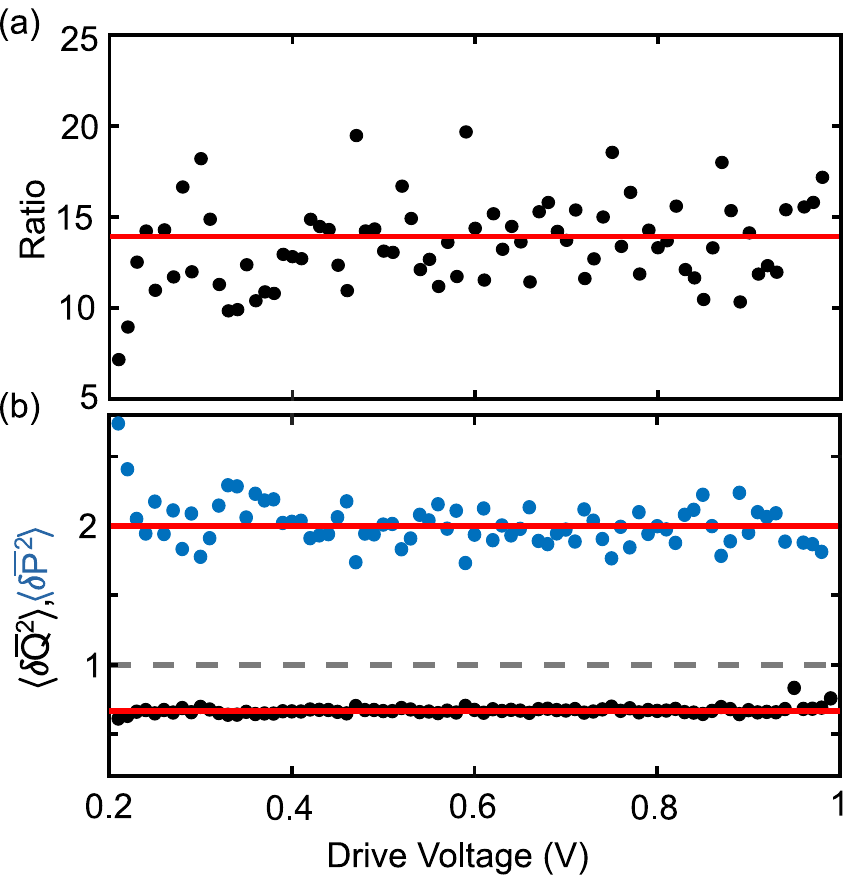}
 	\caption{\textbf{(a)} Ratio of the areas of the satellites peaks obtained from the quadrature data as a function of the drive voltage (black dots). The red solid line corresponds to the theoretical prediction Eq.~(13).
 		\textbf{(b)} Normalized dimensionless variances of the in-phase and quadrature fluctuations $\langle \overline{\delta Q^2}\rangle =  (\omega_0^2/k_BT)\langle \delta Q^2\rangle$  and $\langle \overline{\delta P^2}\rangle =  (\omega_0^2/k_BT)\langle \delta P^2\rangle$  around the stable state of the forced vibration as functions of the drive voltage. Black and blue dots show the in-phase and quadrature values extracted from the experimentally determined satellite area ratio from  (a), whereas the red lines show the corresponding theoretical model of Eq.~(14). The gray dashed line illustrates the thermomechanical fluctuations at the temperature $293$\,K used in the experiment.}
 	\label{fig:figure3_sm}
 \end{figure}

 The same measurements as discussed in Secs. \ref{sec:observations} and \ref{sec:exact_resonance} are repeated and the experimental results are plotted in Fig. \ref{fig:figure2_sm} and Fig. \ref{fig:figure3_sm}.
 Figure \ref{fig:figure2_sm} shows the color-coded response spectra for increasing drive voltage. For sample 2 much stronger drive voltages have to be applied to resolve the satellite peaks. The two distinct peaks, equally spaced from the frequency of the strong drive and with a power-dependent splitting are well resolved and show a good agreement with the theoretical predicted position (red dotted line). In panel (b) a linescan of the amplitude (blue) and quadrature component (black) at $V_d=600$\,mV is plotted. The dip in the quadrature data at the position of the lower frequency satellite peak indicates the amplification of the driving force.
 
 The data taken on sample 2 are analyzed in the same way as described for sample 1 in Sec. \ref{sec:exact_resonance}. Figure \ref{fig:figure3_sm}(a) and (b) show the results for the ratio of the areas of the satellite peaks and the normalized variances of the in-phase and quadrature fluctuations obtained from the quadrature data as a function of the drive voltage. The theoretical model is plotted in both panels as a red solid line and good agreement between theory and experiment is found.
 The averaged squeezing parameter amounts to $0.27 \pm 0.03$ and  is again in very good agreement with the theoretically obtained value of $0.27$. 

\bibliographystyle{apsrev4-1} 	
%

\end{document}